\pdfoutput=1
\documentclass[useAMS,amssymb,usenatbib]{mn2e}
\usepackage{epsfig}
\usepackage{color}
\usepackage{graphicx}
\usepackage{tabularx}
\usepackage{longtable}
\usepackage{hyperref}
\usepackage[normalem]{ulem}

\newcommand{\Ms}{\mathrm{M}_\odot}

\title[Bent by baryons: the low mass galaxy-halo relation]{Bent by
  baryons: the low mass galaxy-halo relation}

\author[Sawala et al.]  {\parbox{\textwidth}{Till
    Sawala$^{1}$\thanks{E-mail: \texttt{till.sawala@durham.ac.uk}},
    Carlos S. Frenk$^1$, Azadeh Fattahi$^2$, Julio F. Navarro$^2$,
    Richard G. Bower$^1$, Robert A. Crain$^3$, Claudio Dalla
    Vecchia$^4$, Michelle Furlong$^1$, Adrian Jenkins$^1$, Ian
    G. McCarthy$^5$, Yan Qu$^1$, Matthieu Schaller$^1$, Joop
    Schaye$^3$ and Tom Theuns $^{1,6}$}\vspace{0.4cm}\\
\parbox{\textwidth}{
$^{1}$Institute for Computational Cosmology, Department of Physics, University of Durham, South Road, Durham DH13LE, UK \\ 
$^{2}$Department of Physics and Astronomy, University of Victoria, 3800 Finnerty Road, Victoria, British Columbia V8P 5C2, Canada\\ 
$^{3}$Leiden Observatory, Leiden University, Postbus  9513, 2300 RA Leiden, The Netherlands \\ 
$^{4}$Instituto de Astrof\'{i}sica de Canarias, C/ V\'{i}a L\'{a}ctea s/n, 38205 La Laguna, Tenerife, Spain \\
$^{5}$Astrophysics Research Institute, Liverpool John Moores University, 146 Brownlow Hill, Liverpool L3 5RF\\
$^{6}$Department of  Physics, University of Antwerp, Campus
Groenenborger, Groenenborgerlaan 171, B-2020 Antwerp, Belgium
 }}

\begin{document}

\date{Accepted 2014 ***. Received 2014 ***; in original
  form 2014}

\pagerange{\pageref{firstpage}--\pageref{lastpage}} \pubyear{2014}

\maketitle

\label{firstpage}

\begin{abstract}
  The relation between galaxies and dark matter halos is of vital
  importance for evaluating theoretical predictions of structure
  formation and galaxy formation physics. We show that the widely used
  method of abundance matching based on dark matter only simulations
  fails at the low mass end because two of its underlying assumptions
  are broken: only a small fraction of low mass ($<10^{9.5}\Ms$) halos
  host a visible galaxy, and halos grow at a lower rate due to the
  effect of baryons. In this regime, reliance on dark matter only
  simulations for abundance matching is neither accurate nor
  self-consistent.  We find that the reported discrepancy between
  observational estimates of the halo masses of dwarf galaxies and the
  values predicted by abundance matching does not point to a failure
  of $\Lambda$CDM, but simply to a failure to account for baryonic
  effects. Our results also imply that the Local Group contains only a
  few hundred observable galaxies in contrast with the thousands of
  faint dwarfs that abundance matching would suggest. We show how
  relations derived from abundance matching can be corrected, so that
  they can be used self-consistently to calibrate models of galaxy
  formation.
\end{abstract}


\begin{figure*}
 \begin{center}
   \vspace{-.2cm}
   \includegraphics*[trim = 0mm 0mm 0mm 0mm, clip, width  =.5\textwidth]{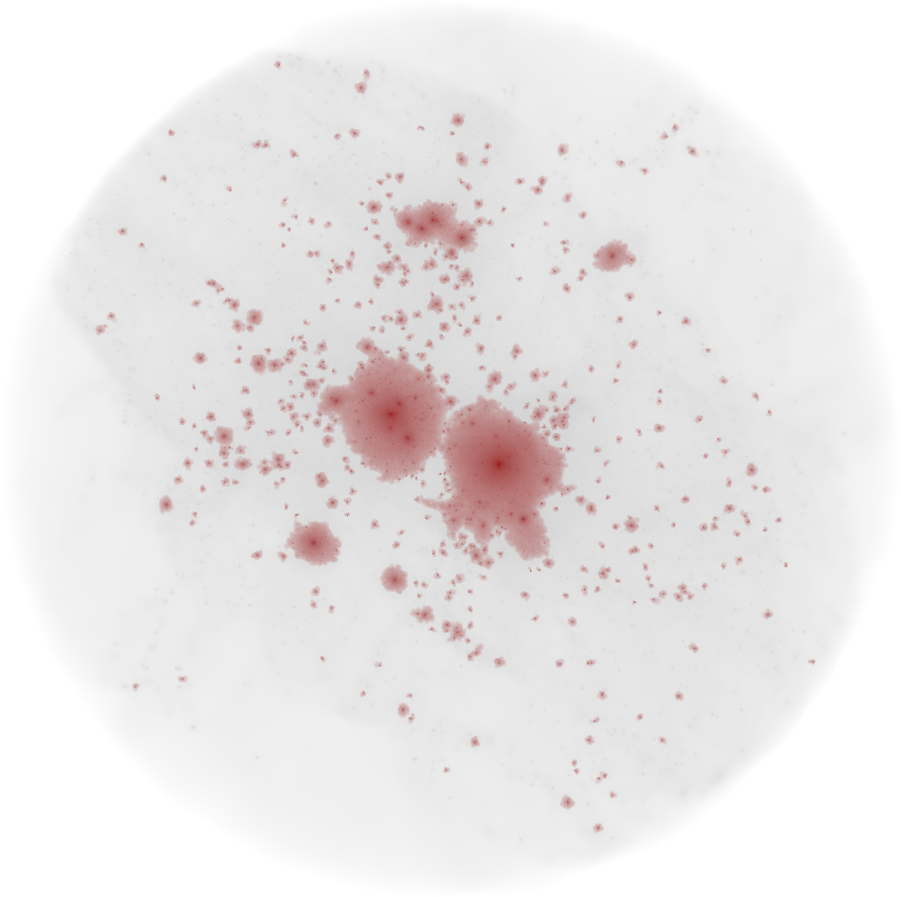}
   \hspace{-.1cm}\includegraphics*[trim = 0mm 0mm 0mm 0mm, clip, width = .5\textwidth]{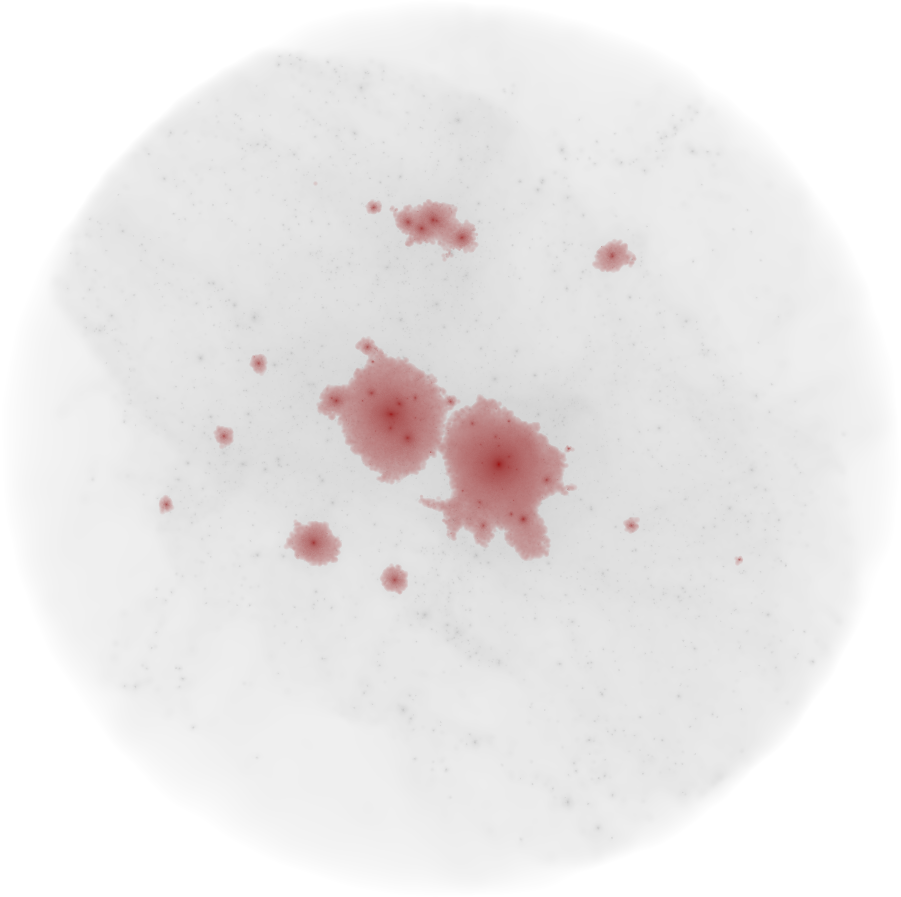} 
 \end{center}
 \caption{Projected density distribution of dark matter within 2 Mpc
   around the simulated Milky Way - M31 barycentre at $z=0$ from one
   of our simulation volumes. Highlighted in red on top of the total
   mass distribution are particles in halos above $5\times10^7\Ms$
   (left panel), and particles in just those halos that contain stars
   (right panel).}
  \label{fig:dark-luminous}
\end{figure*}

\section{Introduction} \label{introduction} In the $\Lambda$CDM
paradigm \citep[e.g.][]{Davis-1985, Frenk-1988}, stars form as gas
cools in collapsed dark matter halos \citep[e.g.][]{White-1978,
  White-Frenk-1991}. The formation of galaxies involves both baryons
and dark matter, but while only gas and stars are directly observable,
numerical simulations of structure formation have largely been limited
to the dark matter component, under the assumption that gravity is the
only relevant force on large scales.

A conceptually simple and yet very powerful method to connect galaxies
and halos in this scenario, which does not require any detailed
knowledge of the complex physics of galaxy formation, is abundance
matching \citep[e.g.][]{Frenk-1988, Yang-2003, Kravtsov-2004,
  Vale-2006, Moster-2009, Guo-2010, Behroozi-2013,
  Moster-2013}. Assuming that a monotonic relation exists between some
observable property of a galaxy (such as stellar mass) and some
property of its dark matter halo (such as total mass), the
relationship between both quantities can be computed from:
\begin{equation}\label{eqn:abundance-matching}
\int_{\rm{a_{h,min}}}^{\rm{a_{h,max}}} N_h(a_h) da_h =
\int_{\rm{a_{\star,min}}}^{\rm{a_{\star,max}}} N_\star(a_\star) da_\star,
\end{equation}
where $N_h(a_h)$ and $N_\star(a_\star)$ are the numbers of halos and
galaxies in the same volume, with properties $a_h$ and $a_\star$,
whose maxima determine the upper limits $\rm{a_{h,max}}$ and
$\rm{a_{\star,max}}$. For any lower limit, $\rm{a_{h,min}}$, a
corresponding lower limit, $\rm{a_{\star,min}}$, is chosen such that
Eq.~\ref{eqn:abundance-matching} is satisfied, and the average
relation $a_\star(a_h)$ is then uniquely determined.

Abundance matching is employed widely to infer quantities such as the
stellar-to-total mass relation \citep[e.g.][]{Frenk-1988, Yang-2003,
  Guo-2010, Moster-2013}. It has also been used to constrain the mass
of the Milky Way's halo from its stellar mass \citep{Guo-2010}, and to
predict the total number of dwarf galaxies in the Local Group from an
N-body simulation \citep{Garrison-Kimmel}.

Generally regarded as assumption-free, abundance matching results are
often interpreted as direct predictions of the underlying cosmological
model. They have also been used as a benchmark for models of galaxy
formation physics, such as semi-analytic models
\citep[e.g.][]{Guo-2011}, and to calibrate hydrodynamic simulations
\citep[e.g.][]{Scannapieco-2012, Munshi-2013}. However, models that
reproduce the abundance matching relation for low-mass galaxies often
require very strong feedback, which can result in an unrealistically
high passive fraction \citep{Fontanot-2009, Weinmann-2012,
  Moster-2013}. Conversely, many hydrodynamical simulations that
produce realistic dwarf galaxies appear to have halo masses
significantly below those inferred by abundance matching
\citep[e.g.][]{Sawala-Matter, Avila-Reese-2011}.

In some cases, the stellar-to-total mass relation derived from
abundance matching can also be compared directly to observations of
individual galaxies. While they agree for galaxies in halos more
massive than $\sim10^{12}\Ms$ \citep{Guo-2010}, discrepancies have
been reported for lower mass halos \citep{Ferrero-2012}. In
particular, dynamical mass estimates derived from stellar kinematics
suggest stellar-to-total mass ratios for individual dwarf galaxies
which are an order of magnitude higher than those inferred from
abundance matching in $\Lambda$CDM.

Recently, we have shown that simulations that model the evolution of
dark matter and baryons as a single fluid subject only to gravity
(henceforth referred to as ``Dark Matter Only'' or DMO simulations) do
not produce the same abundance of halos as hydrodynamic simulations
that include the full baryonic effects. In particular, the collapse
and subsequent expulsion of baryons by feedback processes reduce the
mass of individual halos \citep{Sawala-abundance}, a result that has
since been reproduced and extended to higher masses
\citep{Velliscig-2014, Cui-2014}.

Here we use a new set of high-resolution hydrodynamical simulations of
Local Group volumes to explore how the results of abundance matching
are affected when the effects of baryons and the appearance of dark
halos are included self-consistently. Unlike models that are
calibrated to reproduce an abundance matching relation based on a DMO
simulation {\it before} baryons are taken into account, the
stellar-to-total mass relation produced in our simulations is
consistent with the relation we derive from abundance matching {\it
  after} the effects of baryons are included. Furthermore, this
relation agrees with the dwarf galaxy data, thus demonstrating that
the reported high stellar-to-total mass ratios of these galaxies are
entirely consistent with the $\Lambda$CDM model.

This paper is organised as follows. In Section~\ref{sec:methods}, we
describe the simulations on which our work is based. In
Section~\ref{sec:results}, we describe our results: the fraction of
halos that host galaxies is discussed in
Section~\ref{sec:real-abundance}, the application to abundance
matching in Section~\ref{sec:abundance-matching}, and the implications
in Section~\ref{sec:implications}. We conclude with a summary in
Section~\ref{sec:summary}.

\section{Methods} \label{sec:methods}We use a suite of cosmological
``zoom'' simulations that each contain a pair of galaxies in halos of
$\sim10^{12}\Ms$ in a Local Group environment (see
Fig.~\ref{fig:dark-luminous}). In particular, we require that each
volume contain two halos of $5\times10^{11}-2.5~\times 10^{12}\Ms$
separated by $800 \pm 200$ kpc, approaching with radial velocity of
$0-250$ kms$^{-1}$ and with tangential velocity below $100$ kms$^{-1}$
in an environment with an unperturbed Hubble flow out to 4 Mpc. A
total of 12 volumes were selected from the {\sc Dove} simulation, a
$100^3$Mpc$^3$ N-Body simulation based on the WMAP-7 cosmology. The
cosmological parameters and the linear phases of {\sc Dove}, which are
taken from the public multi-scale Gaussian white noise field {\sc
  Panphasia}, are given in tables 1 and 6 of \cite{Jenkins-2013},
which also describes the method used to make the Local Group zoom
initial conditions. The high resolution initial conditions were set up
using secord-order Lagrangian perturbation theory as described in
\cite{Jenkins-2010}.

The simulations were performed using a pressure-entropy variant
\citep{Hopkins-2012} of the Tree-PM SPH code {\sc P-Gadget3}
\citep{Springel-2005}, described in Dalla Vecchia et al. 2014 (in
prep.). The subgrid physics model is an improved version of the model
used in the {\it OverWhelmingly Large Simulations} project ({\sc
  Owls}; \citealt{Schaye-2010}), that has been developed for the {\it
  Evolution and Assembly of GaLaxies and their Environments project}
({\sc Eagle}, Schaye et al. 2014 in prep., Crain et al. 2014 in
prep.). In brief, the galaxy formation physics model includes
metal-dependent radiative cooling \citep{Wiersma-2009} and
photo-heating in the presence of a UV and X-ray background, and the
cosmic microwave background (CMB). Prior to reionization, net cooling
rates are computed from the CMB and from a UV and X-ray background
that follows the $z=9$ model of \cite{Haardt-2001} with a 1~Ryd
high-energy cutoff. To mimic the temperature boost due to radiative
transfer and non-equilibrium effects over the optically thin limit
assumed in our simulations \citep{Abel-1999}, we inject 2~eV per
Hydrogen and Helium atom. We assume that Hydrogen reionizes
instantaneously at $z=11.5$ \citep{Planck-2013}, while the redshift
dependence of Helium reionization is modelled as a Gaussian centred at
$z=3.5$ \citep{Theuns-2002} with $\sigma(z) = 0.5$. As shown by
\cite{Rollinde-2013}, the resultant evolution of the
temperature-density relation is consistent with measurements of the
intergalactic medium \citep{Schaye-2000}.

Star formation follows \cite{Schaye-2008} with a metallicity-dependent
threshold \citep{Schaye-2004}. The model includes stellar evolution
\citep{Wiersma-2009b} and stochastic thermal supernova feedback
\citep{Dalla-Vecchia-2012}, as well as black-hole growth and AGN
feedback \citep{Rosas-Guevara-2013, Booth-2009, Springel-2005b}.

All simulations were run twice: once with gas and the baryon physics
described above, and once as dark matter only (DMO). To investigate
the regime of Local Group dwarf galaxies with stellar masses below
$10^8\Ms$ and to demonstrate convergence, we use three different
resolutions levels labelled L1, L2 and L3, whose parameters are given
in Table~\ref{table:params}. In the DMO simulations, the dark matter
particle masses are larger by a factor of
$(\Omega_b+\Omega_{DM})/\Omega_{DM}$ relative to the corresponding
hydrodynamic simulations. For this work twelve volumes are analysed at
L3, five volumes at L2, and one volume at L1. In addition, we have
also rerun one volume at L3 and L2 without reionization.

We use a Friends-of-Friends algorithm \citep[FoF, e.g.][]{Davis-1985}
to identify overdense structures (FoF-groups), and the {\sc subfind}
algorithm \citep{Springel-Subfind, Dolag-2009} to identify self-bound
substructures within them. As they represent the objects most directly
associated with individual galaxies, we always refer to the self-bound
substructures as ``halos''. For central and isolated galaxies, the
terms are largely synonymous, but satellite galaxies may share the
same FoF-group while still residing in separate self-bound halos.
\begin{table}
  \caption{Numerical parameters of the simulations} 
\centering 
\begin{tabular}{l l c c c} 
\hline
\hline 
& & \multicolumn{2}{c}{Particle Masses} & Max Softening\\
Label & Type & DM $[\Ms]$ & Gas $[\Ms]$ & [pc] \\ [0.5ex] 
\hline 
L1 & hydro & $5.0\times 10^{4}$ &  $1.0\times 10^{4}$ & 94 \\ 
L1 & DMO & $6.0\times 10^{4}$  & -- & 94 \\
L2 & hydro & $5.9\times 10^{5}$ &  $1.3\times 10^{5}$ & 216 \\ 
L2 & DMO & $7.2\times 10^{5}$  & -- & 216 \\
L3 & hydro & $7.3\times 10^{6}$ &  $1.5\times 10^{6}$ & 500 \\ 
L3 & DMO & $8.8\times 10^{6}$  & -- & 500 \\ [1ex] 
\hline 
\vspace{-.3cm}
\end{tabular}
\label{table:params} 
\end{table}

\section{Results}\label{sec:results}

\subsection{The relevant abundance of
  halos}\label{sec:real-abundance}
The ejection of baryons from low mass halos through heating by the UV
background \citep{Okamoto-2008} and supernova feedback reduces the
growth rate of a halo and, as a result, its mass in the hydrodynamic
simulation is lower than that of the corresponding object in the DMO
simulation. Confirming results of our previous work
\citep{Sawala-abundance} based on the {\sc Gimic} simulations
\citep{Crain-2009}, we find that the average mass of low-mass halos is
reduced by up to $\sim 33 \%$ relative to their counterparts in the
DMO simulation. The universality of this behaviour at the low mass
end, also reproduced by \cite{Munshi-2013}, \cite{Khandai-2014},
\cite{Velliscig-2014}, indicates that it reflects a generic effect of
feedback processes and baryon physics, only weakly dependent on the
particular choice and parameters of the galaxy formation model.

\begin{figure*}
  \begin{center}
    \vspace{-.2cm}
    \hspace{-.2in}\includegraphics*[trim = 5mm 52mm 0mm 10mm, clip, width = .5\textwidth]{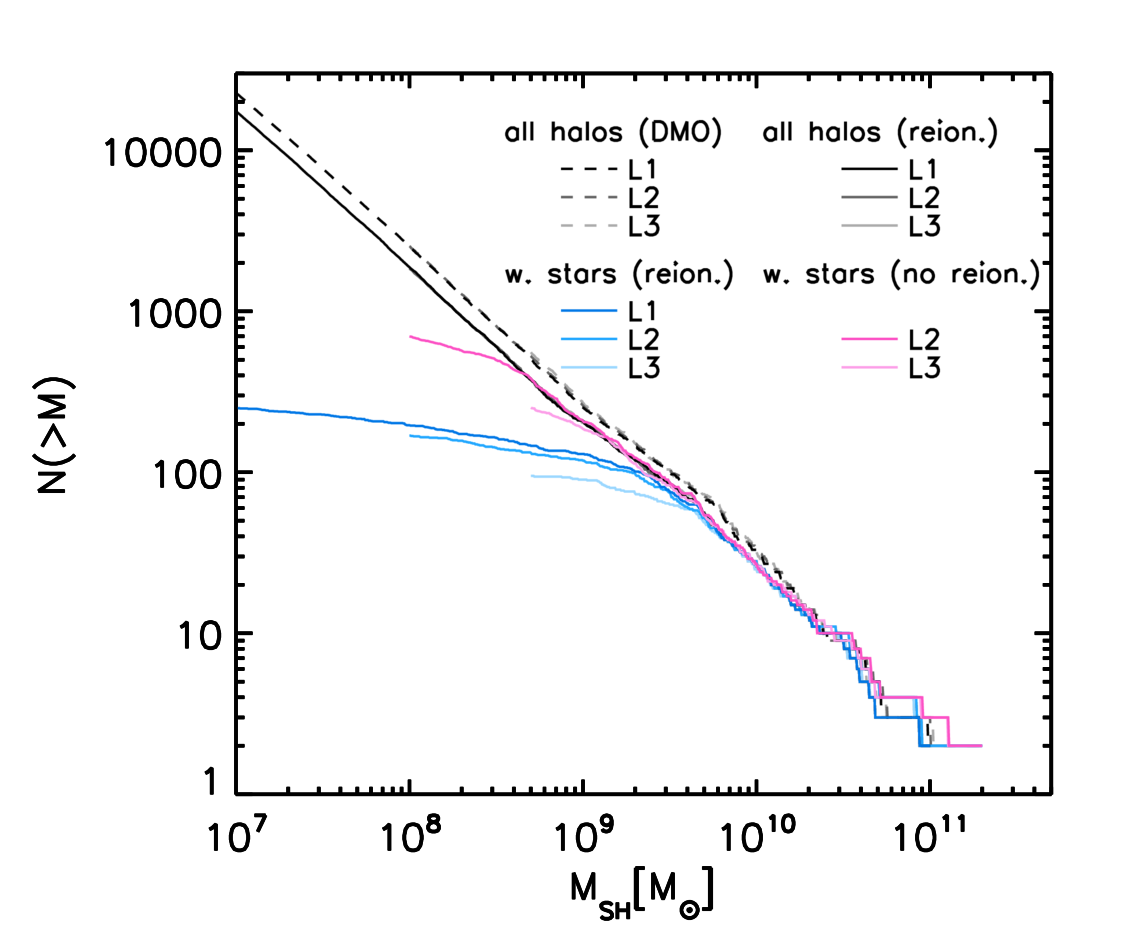}
    \hspace{-.2in}\includegraphics*[trim = 5mm 52mm 0mm 10mm, clip, width = .5\textwidth]{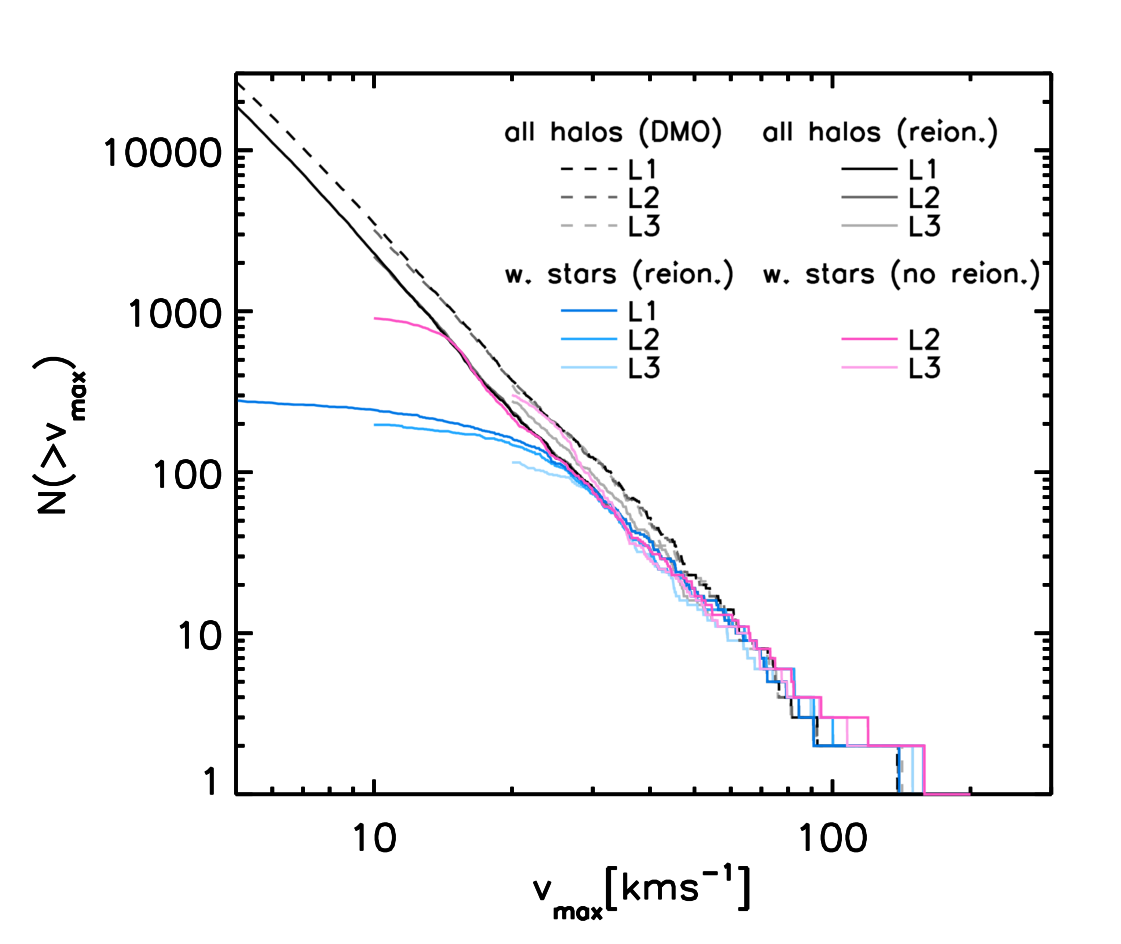} \\
    \hspace{-.2in}\includegraphics*[trim = 5mm 5mm 0mm 25mm, clip, width = .5\textwidth]{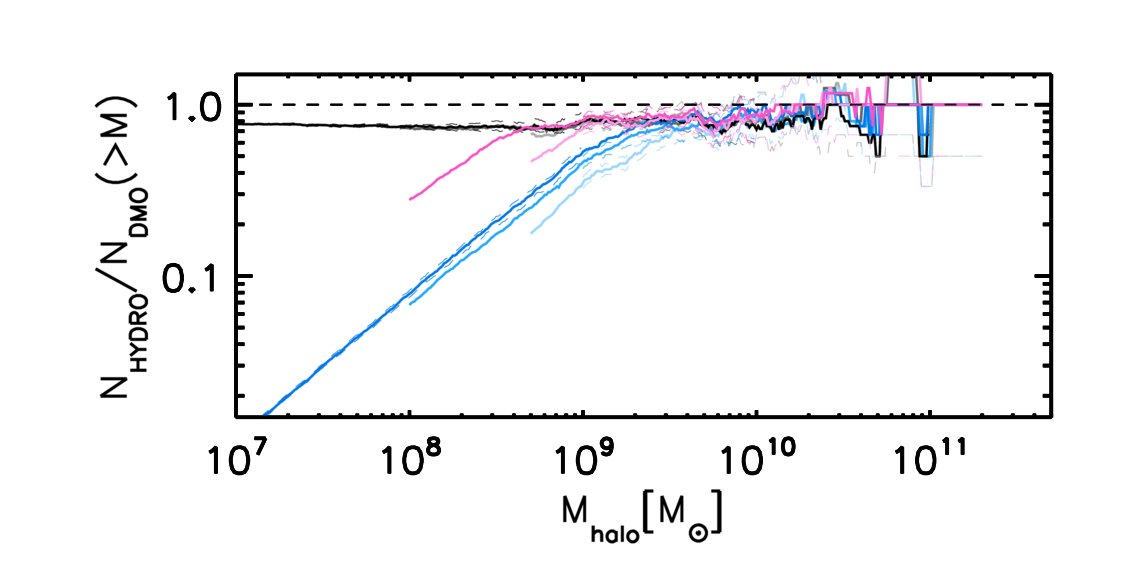}
    \hspace{-.2in}\includegraphics*[trim= 5mm 5mm 0mm 25mm, clip, width = .5\textwidth]{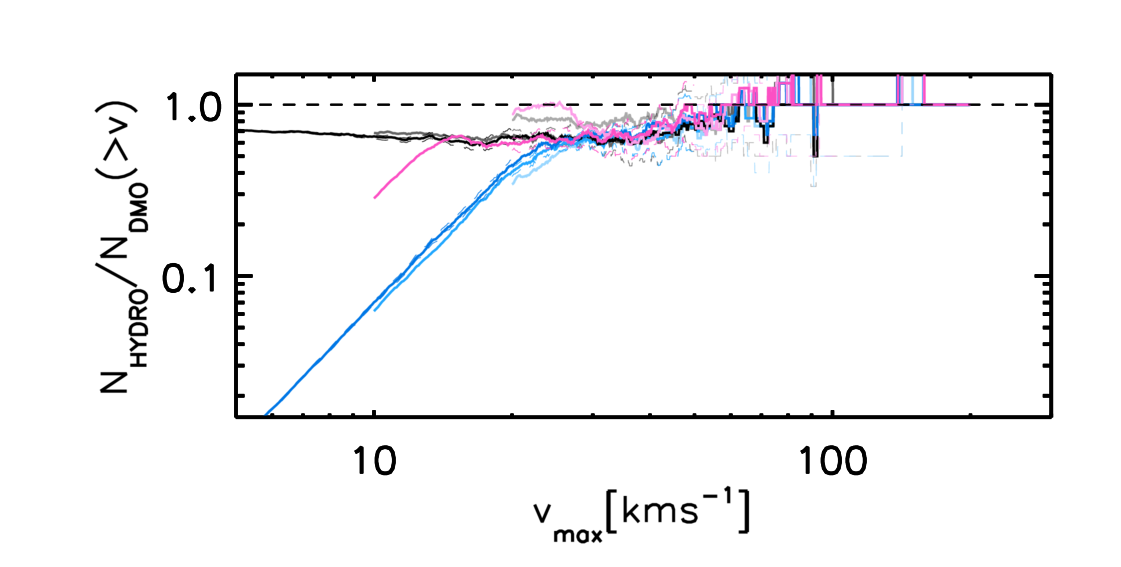}
  \end{center}

  \caption{Top: cumulative halo mass function (left) and maximum
    circular velocity function (right) in one of our simulated LG
    volumes, in DMO simulations (dashed lines) and in hydrodynamic
    simulations (solid lines) at $z=0$. Grey lines show the results
    counting all halos, while coloured lines count only halos that
    contain at least one star particle; blue for the simulation with
    reionization and purple for the simulation without
    reionization. In each case, lightly coloured lines are obtained
    from L3, medium colours from L2, and dark colours from L1. Bottom:
    ratio of the cumulative halo mass function (left) and maximum
    circular velocity function (right), relative to the DMO
    results. Dashed lines also indicate the $1\sigma$ scatter, from
    all twelve volumes at L3 for $\rm{M}>10^{10}\Ms, \rm{v_{max}} >
    50$kms$^{-1}$, five volumes at L2 for $10^9< \rm{M}< 10^{10}\Ms$,
    $30 < \rm{v_{max}}< 50$kms$^{-1}$; and one volume at L1 for
    M$<10^{9}\Ms, \rm{v_{max}} < 20$kms$^{-1}$. While the fraction of
    dark halos decreases significantly from L3 to L2 where the
    threshold stellar mass decreases from $5\times10^6\Ms$ to
    $4.2\times10^5\Ms$, the results are converged for L2 in the
    simulations that include reionization.}  \label{fig:HMF}
\end{figure*}

In the simulations with reionization, we also find that a large
fraction of halos below $10^{9.5}\Ms$ fail to form any
stars. Photoionization has long been recognised as a possible
mechanism to suppress star formation in low mass halos, through a
reduction of cooling \citep[e.g.][]{Efstathiou-1992, Wiersma-2009} and
through the heating and ejection of the interstellar medium
\citep[e.g.][]{Okamoto-2008, Pawlik-2009}. It can be an important
factor in establishing the number of observable dwarf galaxies in the
local universe \citep[e.g.][]{Kauffmann-1993, Bullock-2000,
  Benson-2002, Somerville-2002, Koposov-2009, Okamoto-2009}.

In Fig.~\ref{fig:dark-luminous}, we show the total dark matter
distribution within 2 Mpc from the Local Group barycentre in one of
our simulations. It can be seen that the majority of halos above
$5\times 10^7\Ms$, which are highlighted in the left panel, do not
contain any stars, as highlighted in the right panel. In
Fig.~\ref{fig:HMF} we compare the cumulative halo mass functions (left
column) and halo maximum circular velocity ($\rm{v_{max}}$, the
circular velocity measured at the radius where $\rm{v_{circ}} =
\sqrt{GM(<r) / r}$ is maximal) functions (right column) obtained in
the DMO and hydrodynamic simulations at $z=0$. For the latter, we
further distinguish between the total population of halos, and the
population of halos that contain galaxies. The bottom panels show the
ratios between the cumulative mass- and velocity-functions of the
hydrodynamic and DMO simulations.

In all panels different shades indicate the resolution dependence:
light lines correspond to L3, intermediate to L2 and dark to L1, the
highest resolution simulation. The apparent fraction of dark halos
still decreases between L3 and L2, where the minimum stellar mass
decreases by a factor of 12, from $\sim1.5\times10^6\Ms$ to
$\sim1.3\times10^5\Ms$. However, a further decrease of this minimum
stellar mass by the same factor to $\sim10^4\Ms$ at L1 only has a very
small effect on the total number of galaxies in halos above $10^8
\Ms$, and the corresponding dark fraction. The number of ``galaxies'',
i.e. the fraction of halos in which any stars have formed, has thus
converged at L2.

In the hydrodynamic simulations, the reduced growth rate of halos
below $\sim 10^{10}\Ms$, or $\rm{v_{max}}$ below $\sim 70$kms$^{-1}$,
results in lower masses and a reduced halo abundance. More
importantly, confirming earlier results by \cite[e.g.][]{Crain-2007,
  Okamoto-2008, Okamoto-2009, Pawlik-2009}, we find that reionization
prevents star formation in most halos below $\sim 10^{9.5}\Ms$
($\rm{v_{max}}$ $\sim 25$kms$^{-1}$). We find a continuous decline in
the galaxy fraction, from $\sim100\%$ of all halos at $5\times10^9\Ms$
to $<5\%$ at $10^8\Ms$. Reflecting the fact that the probability of
star formation strongly depends on a halo's formation history and
binding energy, the resulting decline is steeper when expressed in
terms of $\rm{v_{max}}$.

In the simulations without reionization, dark halos only appear for
halo masses below $10^{8.5} \Ms$ or velocities below 12kms$^{-1}$ at
L2. In this (unrealistic) scenario, the total number of galaxies is
not converged in our simulations, and is likely to increase further
with higher resolution. By contrast, the inclusion of reionization
provides an upper limit to the number of galaxies that is independent
of resolution at the level reached in our simulations.

It is also worth noting that while the abundance of objects in the
hydrodynamical simulations with reionization is always at or below the
DMO case in terms of halo mass, it can exceed the DMO case in terms of
$\rm{v_{max}}$ for halos with $\rm{v_{max}}> \sim 100$kms$^{-1}$.  In
these massive halos, cooling is efficient, baryons are largely
processed inside the halo, and adiabatic contraction can lead to an
increase of concentration and $\rm{v_{max}}$, without changing the
total halo mass. We note, however, that as a result of AGN feedback,
the abundance of haloes is reduced up to much higher masses
\citep{Velliscig-2014, Cui-2014}.

\subsection{Abundance matching in the real
  universe}\label{sec:abundance-matching}
While the reduction in average halo mass due to baryons is noticeable,
the steeply declining fraction of halos that contain observable
galaxies has the strongest impact on abundance matching for halos
below $3\times10^{9}\Ms$.

We stress that in order to relate observed galaxies to their dark
matter halos via abundance matching, only those halos that host
galaxies should be taken into account. Defining the total mass
function, $N^\prime_h(m)$, as the mass function of all dark matter
halos irrespective of occupation, and the {\it reduced halo mass
  function}, $N_h(m)$, as the mass function of halos {\it that host
  galaxies}, we can derive the combined effects of the influence of
baryons on halo mass, and the incomplete occupation of halos by
galaxies. Matching $N_h(m)$ as measured in our hydrodynamic simulation
to $N^\prime_h(m)$ as measured in the corresponding DMO simulation,
\begin{equation}\label{eqn:abundance-matching-real}
\int_{\rm{M_{h}}}^{\rm{M_{h,max}}} N_{h}(m) dm = \int_{\rm{M^\prime_{h}}}^{\rm{M^\prime_{h,max}}} N^\prime_{h}(m) dm ,
\end{equation}
we can compute the mass $M_{h}(M^\prime_h)$ of halos containing galaxies in
the hydrodynamic simulation for which the cumulative abundance equals
that of all halos of mass $M^\prime_{h}$ in the DMO simulation.

\begin{figure}
  \begin{center}
    \includegraphics*[trim = 15mm 0mm 0mm 15mm, clip, width = .48\textwidth]{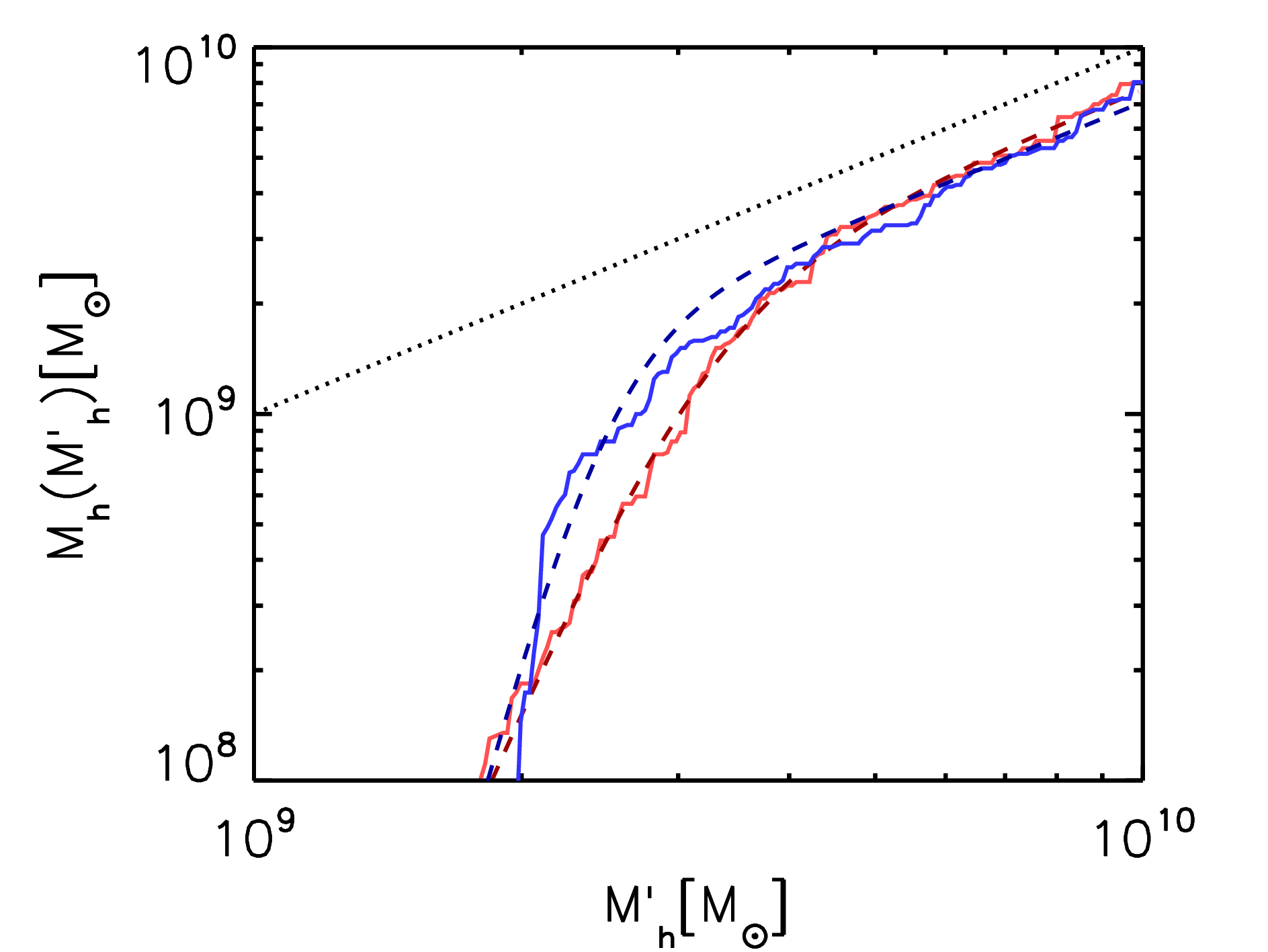}
  \end{center}
  \caption{Mass of halos containing galaxies in the hydrodynamic
    simulation, as a function of the halo mass in the DMO simulation
    corresponding to the same cumulative abundance. Solid curves
    indicate results at $z=0$ (red) and at peak mass (blue), dashed
    curves show the respective fits of Equation~\ref{eqn:mass-fit}
    over the plotted mass interval. For guidance, the dotted black
    line indicates a 1:1 relation, which corresponds to no baryon
    effects and all halos populated by galaxies.}
  \label{fig:equal-abundance}
\end{figure}

As shown in Fig.~\ref{fig:equal-abundance}, for $M^\prime_h =
2\times10^{9} - 10^{10}\Ms$, we find a good fit to the relation
$M_h(M^\prime_h)$ by functions of the form
\begin{equation}\label{eqn:mass-fit}
  M_h(M^\prime_h) = \frac{\alpha M^\prime_h} {1 + (M^\prime_h/M_{0})^\beta},
\end{equation}
where $0 < \alpha < 1$ and $\beta < -1$. The high mass asymptote $ M_h
= \alpha M^\prime_h$ results from the reduction in halo mass by baryon
effects, while the low mass asymptote $M_h = \alpha M_0^\beta
M_h^{\prime1-\beta}$ is caused by the steep decline in the fraction of
luminous halos below $M_{0}$. When measured at $z=0$, we find values
of $\alpha=0.76, M_0 = 3.2\times 10^9\Ms$ and $\beta = -4.7$.

As abundance matching is commonly applied to the peak mass that a halo
attains, rather than to the current mass (which can be reduced by
tidal stripping after infall in the case of satellites), we repeat our
analysis for the same set of halos selected at $z=0$, but using the
peak mass for both $M_h$ and $M^\prime_h$. Here, we find values of
$\alpha=0.71, M_0 = 2.5\times 10^9\Ms$ and $\beta = -8$, but note that
the luminous fraction among halos with peak masses at $\sim10^9\Ms$ in
our simulation is so low that the value of $\beta$ is poorly
constrained.

The relation $M_h(M^\prime_h)$ allows us to obtain an expression for
the stellar-to-total mass relation expected in a universe with baryons
and both luminous and dark halos, given a stellar-to-halo mass
relation previously obtained from abundance matching, where a DMO
simulation and a complete occupation of halos by galaxies had been
assumed:
\begin{equation} \label{eqn:mass-at-equal}
 \left.\frac{m_\star}{m_{H}}\right\vert(M_{h}) =  \left.\frac{m_\star}{m_{H}}\right\vert_{DMO} (M^\prime_h),
\end{equation}
under the assumption that there is no significant change in the
correlation between average halo mass and galaxy mass.

Since those halos that are predicted not to host any galaxy are
excluded from the {\it reduced halo} mass function, galaxies are now
assigned to halos of lower masses than they would have been in the DMO
simulation, increasing their stellar-to-total mass ratios. By analogy,
if galaxies exist that do not belong to any dark matter halos
(e.g. ``tidal dwarf galaxies''), we would require a {\it reduced
  galaxy} stellar mass function, with the effect of decreasing the
average stellar-to-total mass ratios. However, such objects are not
found in our simulations.

\begin{figure}
  \begin{center}
\includegraphics*[trim = 12mm 148mm 40mm 35mm, clip, width = .5\textwidth]{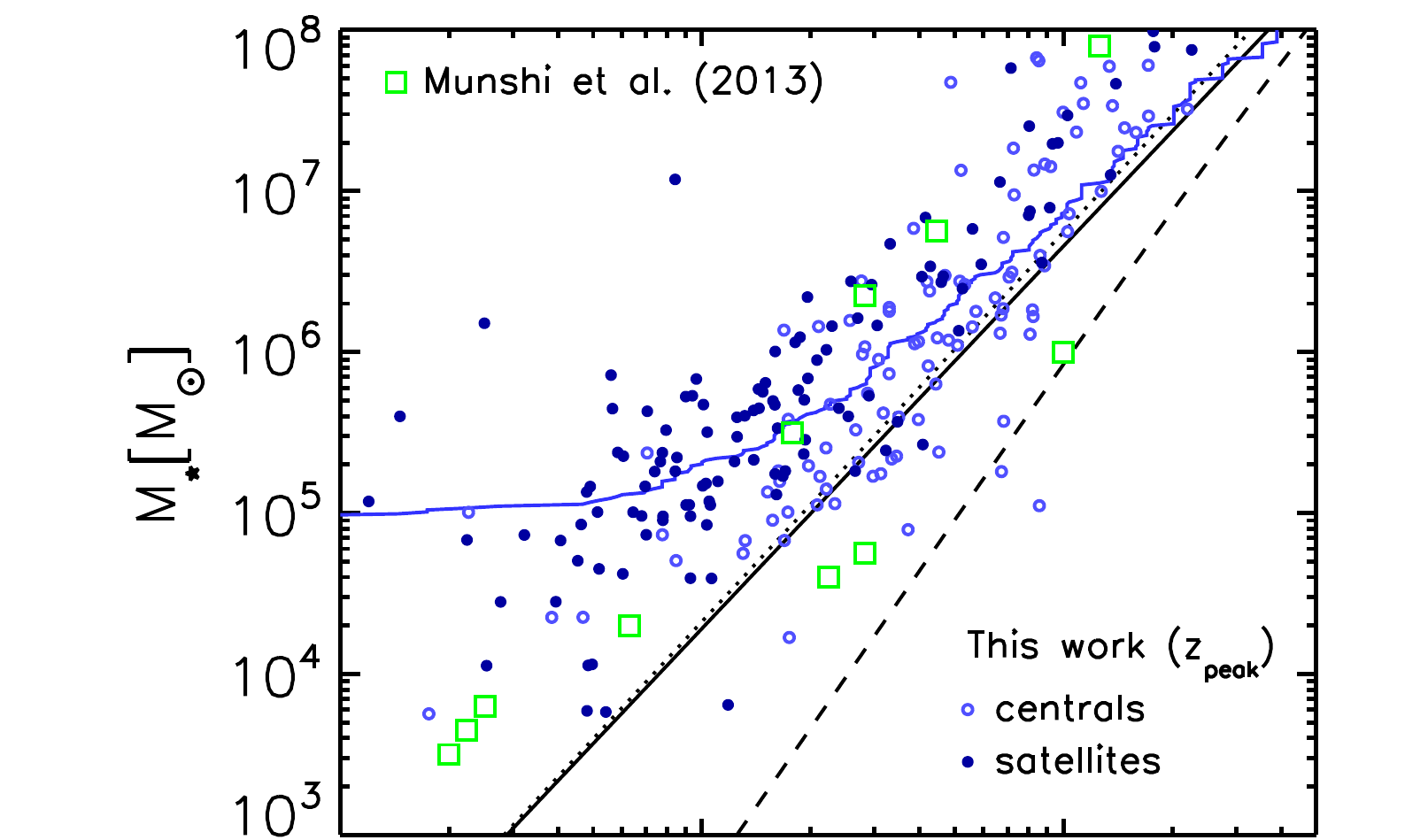}
\includegraphics*[trim = 12mm 126mm 40mm 35mm, clip, width = .5\textwidth]{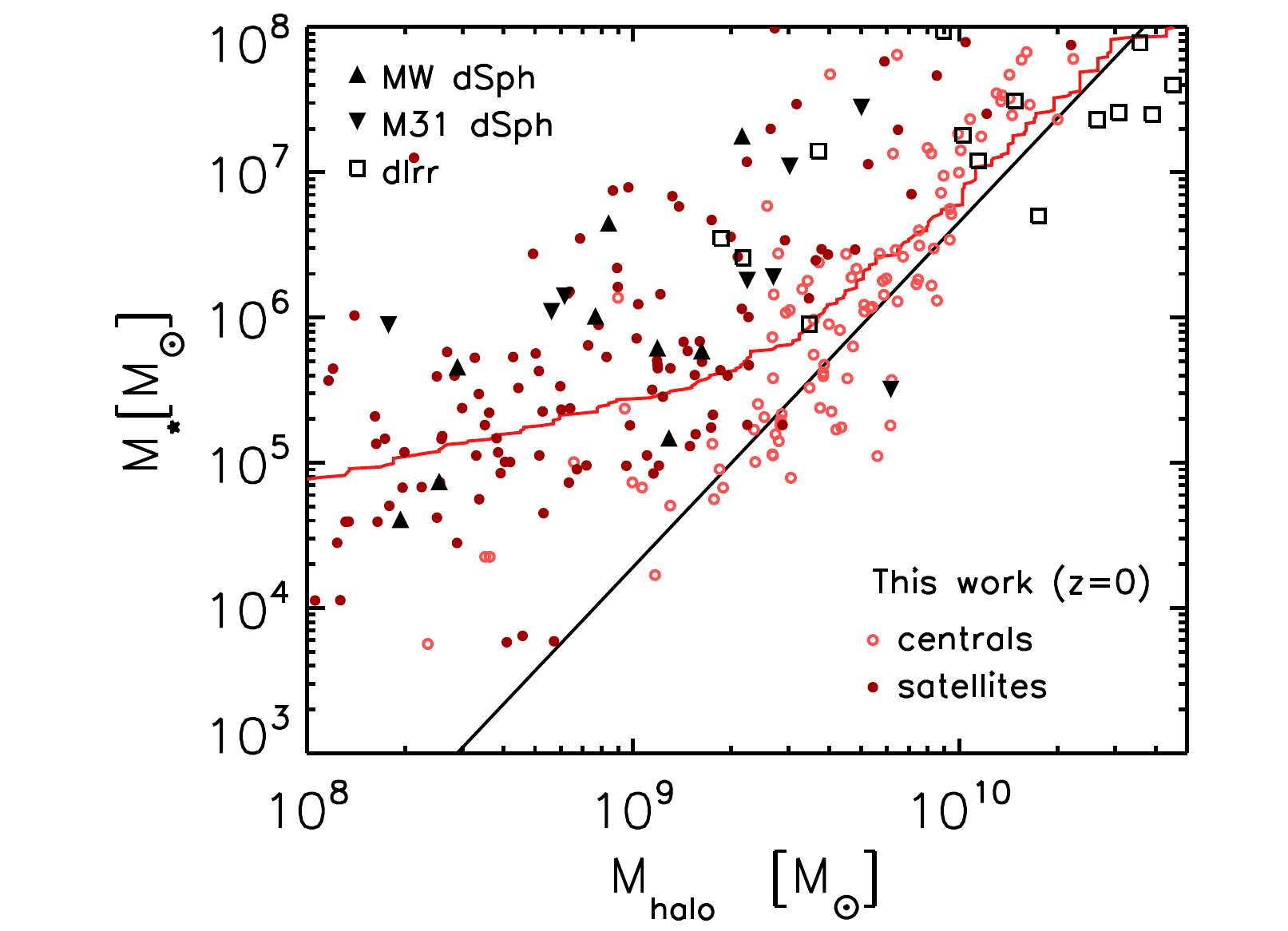}

 \end{center}
 \caption{Stellar-to-total mass relation for galaxies as a function of
   peak halo mass (top) and of halo mass at $z=0$ (bottom). The black
   lines show abundance matching results from DMO simulations by
   \citet[dashed]{Guo-2010}, \citet[dotted]{Wang-2010}, and
   \citet[solid, reproduced in both panels]{Moster-2013}. The blue
   solid line (top) and red solid line (bottom) show the results of
   \citet{Moster-2013} after our correction for the effect of baryons
   and dark halos. In both panels, circles show the results for
   central galaxies (open circles) and satellite galaxies (filled
   circles) as measured in our simulations. In the top panel, green
   squares denote results of hydrodynamic simulations by
   \citet{Munshi-2013}, which are tuned to reproduce the original
   abundance matching relation. In the bottom panel, we compare our
   results at $z=0$ to observations of individual M31 and MW dwarf
   spheroidal galaxies (triangles, \citet{Misgeld-2011, Woo-2008,
     Penarrubia-2008, Tollerud-2012}), and dwarf irregular galaxies
   (squares, \citet{Ferrero-2012, McGaugh-2005, Oh-2011, Stark-2010,
     Cote-2000}).}
  \label{fig:abundance-matching}
\end{figure}

In Fig.~\ref{fig:abundance-matching}, we show the resulting change in
the stellar-to-total mass relations, with halo masses measured at peak
mass (top), or at $z=0$ (bottom). The black lines show the results
based on standard abundance matching of DMO simulations by
\citet[dashed]{Guo-2010}, \citet[dotted]{Wang-2010}, and
\citet[solid]{Moster-2013}, while the blue line (top panel) and red
line (bottom panel) show the relation of \citet{Moster-2013} after our
correction for baryonic effects and the presence of dark halos.

In both panels, it can be seen that the typical stellar mass of
galaxies in halos of $10^9\Ms$ is increased by an order of magnitude,
and in halos of $3\times10^8\Ms$, it is increased by almost two orders
of magnitude. This change in the stellar-to-total mass ratio can be
greater than the differences between individual abundance matching
models that are shown in the top panel and which account for
differences in the stellar mass function, cosmology, assumed scatter,
and treatment of satellites. In other words, abundance matching
applied to DMO simulations that do not account for baryons fail to
constrain any of the other parameters in the dwarf galaxy regime.

It can also be seen from Fig.~\ref{fig:abundance-matching} that the
faint end tail of the stellar-to-total mass relation deviates from the
single power-law form which was originally introduced by
\cite{Yang-2003} and that is assumed in most abundance matching
models.

\subsection{Self-consistent abundance matching}\label{sec:implications}
The stellar and total masses measured for galaxies and halos in our
simulation at L1 are indicated by circles in both panels in
Fig.~\ref{fig:abundance-matching}. Both at $z=0$ and at peak mass,
these values agree well with the stellar-to-total mass relations
derived from abundance matching when baryons are taken into account,
as described in Section~\ref{sec:abundance-matching}, but not with the
original DMO abundance matching results.

The bottom panel of Fig.~\ref{fig:abundance-matching} also includes a
comparison to observational data from Milky Way and M31 satellite
galaxies, and dwarf irregular galaxies. For the MW satellites we use
virial masses estimated by \cite{Penarrubia-2008}, while for M31
satellites we convert $\rm{v_{max}}$ estimates by \cite{Tollerud-2012}
using the relation for satellites given by \cite{Klypin-2011}. For the
dwarf irregulars, we show $m_{200}$ for the best-fitting NFW profile
as measured by \cite{Ferrero-2012}, limiting the sample to galaxies
where the velocity profile is measured to at least $r_{out} = 200pc$.

For consistency with the observations, the halo masses at $z=0$ of
individual objects in the simulation are derived using the measured
maximum circular velocity converted to mass following
\cite{Klypin-2011}. While these observed mass estimates still involve
a high degree of uncertainty, our simulation and the corrected
abundance matching relation are both in good agreement with the
observational data. The discrepancy between abundance matching and the
measurements of individual galaxies reported by \cite{Ferrero-2012} is
resolved once the abundance of halos is corrected.

It should perhaps also be noted that abundance matching based on DMO
simulations is not entirely self-consistent: a generic prediction
obtained through abundance matching is a baryon fraction of dwarf
galaxies far below the universal value, which already indicates a
change in total halo mass not included in the underlying DMO
simulation.

Also shown by green symbols in the top panel of
Fig.~\ref{fig:abundance-matching} are the results of hydrodynamical
simulations by \cite{Munshi-2013}, which ``successfully'' reproduce
the abundance matching relation obtained from a DMO simulation. Note
here that at $2\times10^8\Ms$, fewer than one in ten halos host a
galaxy in our simulations, whereas the relation reproduce by
\cite{Munshi-2013} assumes all halos to be populated. As the
stellar-to-total mass ratios diverge at low masses from those obtained
in our simulation (blue symbols), the assumed models of galaxy
formation physics must differ. In particular, in order to satisfy the
constraints imposed by abundance matching based on DMO simulations,
\cite{Munshi-2013} impose a very low star formation efficiency due to
processes like supernova feedback. Our results suggest that a reliance
on abundance matching based on DMO simulations can lead to artificial
conclusions for both cosmology and astrophysics.

It is also worth noting that our simulations with reionization only
predict a few hundred dwarf galaxies within a radius of $2.5$Mpc of
the Local Group, in contradiction with recent results obtained from
DMO simulations and abundance matching by
\cite{Garrison-Kimmel}. These authors predict thousands of as-yet
undiscovered faint dwarf galaxies, whereas our simulations predict
that the Local Group should contain thousands of dark halos, only a
small fraction of which contain galaxies.

\section{Summary}\label{sec:summary}
The commonly used method of abundance matching between N-body DMO
simulations and the observed stellar mass function is not assumption
free: it relies on the implicit assumption that structure formation
can be represented by DMO simulations and that every halo hosts a
galaxy. Using a set of high resolution simulations of Local Group
volumes with and without baryons, we have shown that, at the low mass
end, both of these assumptions are broken. Because of cosmic
reionization, most halos below $3\times 10^9\Ms$ do not contain an
observable galaxy. In this regime, the median stellar-to-total mass
relation inferred directly from abundance matching only applies to
(mostly unobservable) halos, but has no direct application to
observable galaxies.

Our simulations assume reionization in the optically thin limit
without self-shielding, and radiative cooling which does not include
molecular cooling at low temperatures. While these limitations may
influence the impact of reionization on star formation rates in
existing galaxies, the proto-galaxies that are prevented from star
formation in our simulations reach neither the gas densities required
for self-shielding nor the metallicities required for molecular
cooling after reionization. Of course, our simulations cannot resolve
the formation of Pop-III stars from primordial gas. Our results also
assume that reionization is uniform and local variations might change
the impact for individual galaxies, but probably not enough to affect
our results substantially.

By equating the cumulative abundances of objects from hydrodynamical
and DMO simulations, we have shown how abundance matching results can
be modified to account for baryonic effects. We stress that a reliance
on the results from abundance matching obtained through dark matter
only simulations can have serious consequences: it can lead to
erroneous conclusions about cosmology, and, when used as a benchmark
for hydrodynamic simulations or semi-analytical models, it can also
lead to false conclusions about galaxy formation physics. Baryon
effects have to be taken into account self-consistently for abundance
matching to give a meaningful interpretation of the connection between
galaxies and halos.

\section*{Acknowledgements}We would like to thank I.~Ferrero for
providing the dIrr data points included in
Fig.~\ref{fig:abundance-matching}. This work was supported by the
Science and Technology Facilities Council [grant number ST/F001166/1
and RF040218], the European Research Council under the European
Union's Seventh Framework Programme (FP7/2007-2013) / ERC Grant
agreement 278594-GasAroundGalaxies, the National Science Foundation
under Grant No. PHYS-1066293, the Interuniversity Attraction Poles
Programme of the Belgian Science Policy Office [AP P7/08 CHARM] and
the hospitality of the Aspen Center for Physics. M.~F. and
T.~S. acknowledge the Marie-Curie ITN CosmoComp, and
M.~F. acknowledges the ERC Advanced Grant programme
Dustygal. C.~S.~F. acknowledges an ERC Advanced Investigator Grant
COSMIWAY. This work used the DiRAC Data Centric system at Durham
University, operated by the Institute for Computational Cosmology on
behalf of the STFC DiRAC HPC Facility (www.dirac.ac.uk), and resources
provided by WestGrid (www.westgrid.ca) and Compute Canada / Calcul
Canada (www.computecanada.ca). The DiRAC system is funded by BIS
National E-infrastructure capital grant ST/K00042X/1, STFC capital
grant ST/H008519/1, STFC DiRAC Operations grant ST/K003267/1, and
Durham University. DiRAC is part of the National
E-Infrastructure. \bibliographystyle{mn2e} \bibliography{gimic}

\label{lastpage}

\end{document}